\documentstyle[11pt,epsf]{article}
\def\ba{\begin{eqnarray}\samepage}
\def\ea{\end{eqnarray}}
\input amssym.def
\input amssym.tex

\def\double{\Bbb}

\newcommand{\po}{Poincar\'{e} }

\def\zz{Z}

\def\zz{{\double Z}}

\newcommand{\siot}{$\mbox{sio}_2$\ }

\newcommand{\diff}{\partial}

\newcommand{\be}{\begin{equation}}
\newcommand{\ee}{\end{equation}}

\newcommand{\ben}{\begin{eqnarray}\displaystyle}
\newcommand{\een}{\end{eqnarray}}

\def\Bbb#1{\fam\msbfam#1}

\def\beq{\begin{equation}}
\def\eeq{\end{equation}}

\begin{document}
\title{An  approach to F-theory}

\author{S. F.  Hewson\footnote{sfh10@damtp.cam.ac.uk}\\ 
Department of Applied Mathematics and Theoretical
Physics\\Silver Street, Cambridge, CB3 9EW\\England}

\maketitle

\begin{abstract}
We consider BPS configurations in theories with two timelike
directions from the perspective of the supersymmetry algebra. We show
that whereas a BPS state in a
theory with one timelike variable must have positive energy,  in
a theory with two times  any
BPS state must have positive angular momentum 
in the timelike plane, in that $Z_{0\tilde{0}}>0$, where $0$ and
$\tilde{0}$ are the two timelike directions. We consider some
generic  BPS 
solutions of theories with two timelike directions, and  then
specialise to the study of the (10,2) dimensional 
superalgebra 
for which the spinor operators generate 2-forms and 6-forms. We argue
that the BPS configurations of this algebra relate to F-theory in the
same way that the BPS configurations of the eleven dimensional
supersymmetry algebra relate to  M-theory. We show that the twelve
dimensional theory is one of  fundamental 3-branes and
7-branes, along with their dual partners. We then formulate the new
intersection rules for these objects. Upon reduction of this system we find the
algebraic description of the IIB-branes and the M-branes.  Given these
correspondences we may begin an algebraic study of F-theory. 
\end{abstract}

\section{Introduction}
A long standing problem in string theory has been the explanation of
the self S-duality of
the type IIB superstring. F-theory \cite{Vaf:ftheory} provides a
higher dimensional  mechanism which suggests a solution to this
problem.  F-theory is defined  such that given a bundle manifold
${\cal{M}}$ which is a 
$T^2$ fibre over a base space ${\cal{B}}$ 
\be
\mbox{F on }{\cal{M}}\equiv \mbox{IIB on }{\cal{B}}\,.
\ee
By allowing the internal moduli of the torus to vary over the base, the
$SL(2,\zz)$ symmetry of the IIB theory is explained. 
By necessity, F-theory   is a twelve dimensional structure and
involves in some way two
timelike directions. This has created an interest in theories with 
dimensions and signatures  beyond those common to supergravity
\cite{HewPer:22brane,Hew:generalised,KutMar:newprinciples,Tse:12dwave,Bar:susy,NisSez:,RudSez:,KhvKhvLuPop:,Nis:2+2,BarKou:2times,RudSezSun:,Tow:12d}.

  In this paper we adopt the
attitude that some local theory exists in (10,2) dimensions, in the
same spirit as  M-theory. We think of M-theory as being some quantum structure
 which reduces to the superstring theories and the eleven dimensional
supergravity theory in different limits.  Many of the properties of
M-theory in the extremal limits may be inferred from an  investigation
of the different types 
of solution to the low energy supergravity theories.
In a supergravity theory, there
exist local 
supersymmetry transformations relating the bosonic and fermionic
fields to each other. In such a theory, however, the behaviour of the
long range fields 
is governed by the superalgebra structure. A study of the algebra thus
provides us with many constraints on the global properties of the 
local theory
 \cite{Tow:fourlectures}. Given these restrictions we may
infer properties about the 
underlying  M-theory. As an example, by studying the superalgebra it
was deduced that one could define the  Green-Schwarz superstring solution
in ten
dimensional flat space \cite{GreenSchwarzWitten:}. It was realised that
this ought to 
imply the 
existence of a string-like solution to the corresponding supergravity
theory, which was later written down explicitly \cite{DabGibHarRui:}.
In a similar way, a membrane-like solution to the eleven dimensional
supergravity equations was constructed \cite{DufSte:membrane},
corresponding to the Green-Schwarz membrane which exists in a flat
 superspace in eleven dimensional spacetime. 
This basic process
has now become much more
elaborate, and the supergravity interpretation of a very general class
of solutions to the rigid
supersymmetry theories are now well understood 
\cite{PapTow:,Gau:intersecting}. 
Although simple, this approach is
very effective and gives us a handle on the possible types of object,
such as 2-branes and 5-branes, 
which appear in
the different limits of M-theory. We now wish to ask the question: Which
objects appear in F-theory? Very little is certain  because there is no
known 
low energy theory to which we can turn for guidance. We do,
however, have at our disposal 
candidate  superalgebras. 
We therefore make a start by studying the
brane structure of 
superalgebras in spacetimes of signature (10,2), which we shall
loosely call F-theory. By analogy with 
results from supergravity it is to be
expected that the objects in a theory of rigid supersymmetry branes
will  be in 1-1 correspondence with those in the  local version of F-theory.

\subsection{Two-timing theories}
 Although it is
currently unclear as to whether or 
not the additional timelike dimension in F-theory  is  to be treated
as physical, in the 
same sense as the extra dimensions in string theory correspond to
extra dimensions in spacetime, or 
merely as some auxiliary variable,  there
is potentially much to be learned from the study of theories with two
times. Although there are many suggestions as to the nature of such
theories, in our opinion the major question which has yet to
be addressed is 
thus: Which theory {\it{should}} we be studying? There are obviously
many ways in 
which {\it some} theory with two timelike directions may be
constructed, but how may we be sure that it corresponds to the one
relevant to physics? One of the more
sensible places to begin with is  the 
known supergravities.   
However, there are   many 
ways in which Minkowski signature supergravity theories may be related to 
theories with different spacetime signature. The 
 most obvious method 
is to consider simply the low energy bosonic sectors of the known
theories and their proposed generalisations,  imposing the requirement that
they match up when linked via a dimensional reduction or duality
transformation.
This approach 
 has been
discussed in various contexts 
\cite{Tse:12dwave,KhvKhvLuPop:,Kar:}. As regards the full
underlying {\it supersymmetric} theories, however,  these procedures
are potentially misleading:  
A reason for this is that the Einstein term in the low energy
effective action arises because the anticommutator of two
supersymmetry generators produces a translation. In the local theory
this creates the diffeomorphisms which allow us to define an Einstein
theory of gravity. Unfortunately there is no guarantee in a
supersymmetry theory 
with two times that the translation generators  appear in the
algebra in a consistent way. The reason for this apparent discrepancy
is that the symmetry  properties of products of gamma-matrices 
are related to the
signature of the spacetime, whereas the number of spin degrees of
freedom grows rapidly with increasing dimension
\cite{Hew:generalised}. With few exceptions
\cite{ChaEmp:,Nis:2+2,Wan:12dsupergravity}, the entire 
supergravity structure which we are used to dealing with is based on
the understanding that there is one timelike direction, and a certain
amount of care must consequently be taken when discussing possible `low energy
supergravities' in a  space with two times. Indeed, even the meaning of
the term `low energy' must be carefully discussed, as an absence of
translations leads to the lack of a simple notion of energy itself!

Notwithstanding the aforementioned complications, the algebraic study
of BPS
configurations allows us to deduce with confidence some 
properties of the F-theory spectrum.  Once the
concrete two-timing supersymmetric  BPS  solutions are known, it is
 possible to discuss the 
relationships to 
Minkowski signature theories.  

\section{BPS configurations in general supersymmetric theories}

One of the most useful concepts of supergravity, or indeed any low
energy limit of a theory which is only understood perturbatively,  is
that of 
the BPS state: From the BPS configuration, which is a special solution
to the effective theory at weak coupling, we can deduce
information 
about the corresponding quantum theory at strong coupling. The reason
for this is 
that the BPS states lie in  short representations of the supersymmetric
theory, and as such should remain so even as the coupling is
switched on \cite{OliWit:}. 
These states are central to the study of Minkowski signature theories,
and if there is a physical link between M-theory and F-theory then it
should be provided by the BPS states in each. We therefore discuss BPS
conditions in a general $T=2$ supersymmetric theory and then specialise to
F-theory. 

\subsection{Supersymmetry algebras}

The supersymmetry algebra is the graded generalisation of the isometry
algebra of the tangent space of a spacetime manifold.  A Lorentz 
supersymmetric 
theory is one such that  the isometries are given by the Lorentz
algebra, which is  
generated by the 
Lorentz rotations $M_{\mu\nu}$,  
supplemented by 
an anticommutator $\{Q^\alpha,Q^\beta\}$ and a commutator
$[Q^\alpha,M_{\mu\nu}]$, 
where $Q$ is a spinorial  
 operator. In addition, other
central terms $Z_{\mu_1\dots\mu_n}$ may appear in the algebra
\cite{Hew:generalised,Bar:s,Tow:democracy}. This leads to the rich
brane structure of 
M-theory \cite{Tow:fourlectures}.

Generically, a quantum vacuum of  a supersymmetric theory is defined to be a
state $|0\rangle$  in an Hilbert space which is annihilated by all the fields
\be\label{vacuum}
Q^\alpha|0\rangle=M_{\mu\nu}|0\rangle
=Z_{\mu_1\dots\mu_n}|0\rangle=0\,.
\ee
A general quantum state $|\psi\rangle$   is not projected to
zero by the fields, and  is labelled
by the eigenvalues of the various  operators.
In addition to $|0\rangle$ and $|\psi\rangle$,  due to the
anticommuting nature of the fermionic 
spinor operators, we may define the intermediate class of BPS
states. These are states which are annihilated by some linear
combination of the spinor generators,  which is equivalent to the
condition that  
\be\label{bpsequation}
\det\left(\{Q^\alpha,Q^\beta\}\right)=0\,.
\ee
These solutions preserve some fraction of the supersymmetry equal to
the ratio of the number of zero eigenvalues of the anticommutator to
the dimensionality of the spin space, and as a result lie in  short
representations of the general supersymmetry algebra.
This result holds quite 
independently of the particular details  of the supersymmetry
algebra,  and allows us to
simply generalise to the situations with two times.
We shall say that a BPS state is given by any
background which is a 
consistent solution to (\ref{bpsequation}). 
By consistent we mean that the generators of the solution must satisfy
the appropriate underlying super Lie algebra. The important terms to
consider are 
the anticommutators of the spinor generators with themselves, which
most generally 
take the form
\be\label{ac}
\{Q^\alpha,Q^\beta\}=\sum_{n=1}^{[D/2]}
\left(C\Gamma^{\mu_1\dots\mu_n}\right)^{\alpha\beta}Z_{\mu_1\dots\mu_n}\,.
\ee
The sum may be taken to  terminate at the integer  part of half
the dimension by employing the \po duals of the higher spin $Z$ fields,
if we ignore topological complications. 
We take the spinor $Q$ to be Majorana,  in which case the left hand
side of  (\ref{ac}) is clearly a real symmetric matrix, which is {\it
positive definite} since we choose to work in a Hilbert
space\footnote{In a Pontryagin space, which is an indefinite metric
Hilbert space, this restriction is waived and we may define more
general supersymmetry theories. This, however, is at the expense of
introducing additional complications in the quantisation
procedure.}. As such, the matrix is diagonalisable and has zero or
positive 
eigenvalues. The components $Z_{\mu_1\dots\mu_n}$ must be chosen so
as to make the right hand side of the equation symmetric, with
non-negative eigenvalues; in particular this means
that the trace of the right hand side must be positive.

\subsection{$T=1, D>2$}
For the signature $(D-1,1)$ Clifford algebra, the charge conjugation
matrix used to raise and lower spinor indices may be taken to be
$C=\Gamma^0$, the timelike gamma-matrix. In this case we find that the 
$Tr\left(C\Gamma^{\mu_1\dots\mu_n}\right)$ are zero except for
$Tr(C\Gamma^0)=Tr(1)=D$. To
see this note that  
$C\Gamma^{\mu_1\dots\mu_n}$ is an anti-symmetrised product of less than
$D$ gamma matrices. The trace of an even number of anticommuting
matrices is always zero, due to the cyclic property of $Tr$. If we
have an odd number of anticommuting matrices, then we insert
$1=(\Gamma^{(s)})^2$ into the trace, where $\Gamma^{(s)}$ is a
spacelike gamma matrix not 
in the product; we may always do this because the integer $n$ of the sum in
(\ref{ac}) only runs up to $[D/2]$. 
This means that the trace of the right hand side of (\ref{ac}) is zero
unless $Z_0\equiv P_0\neq 0$. Since we must sum over symmetric
$\left(C\Gamma^{\mu_1\dots\mu_n}\right)^{\alpha\beta}$, this implies
the existence of 
negative eigenvalues, which is a contradiction. 
We thus arrive at the positive energy condition
\begin{itemize}
\item
A BPS state in a $T=1, D>2$ theory must have  $P_0>0$.
\end{itemize}
This is a pleasing result, since it tells us from  algebraic
considerations that {\it any} brane configuration of a general low energy
supersymmetric theory must have non-zero energy. In addition, the role
of the momentum is uniquely singled out from the set of possible
central charges $Z^{\mu_1\dots\mu_d}$.

\subsection{$T=2, D\geq 4$}

For a theory with two times, the crucial point to notice is that the
charge conjugation matrix is given by  the product of
the two timelike gamma matrices, $C=\Gamma^{0}\Gamma^{\tilde{0}}$
\cite{Nie:supergravity}. In 
this case, the only term on the right hand side of (\ref{ac})
 with non-zero trace is
$C\Gamma^{0\tilde{0}}M_{0\tilde{0}}$. By a similar
argument to that for the $T=1$ case, we find that
\begin{itemize}
\item
A BPS state in a $T=2, D\geq 4$ theory  must  have $Z_{0\tilde{0}}>0$. 
\end{itemize}
In theories with two times, therefore,  
 a special role is given to the
$Z_{\mu\nu}$ terms, and the other $p$-form charges have no particular
constraints (except that the positivity condition must be
satisfied). This is an important point:  
In two time theories we are quite generally allowed
configurations which have zero  $P_0$; the only  requirement is
that the angular momentum in the timelike plane,  $Z_{0\tilde{0}}$, is
positive. Hints of this  type of behaviour have occured in the
 literature, in the context of  the Two-Timing Monopole
 \cite{ChaEmp:}: This is the analogue of the Kaluza-Klein monopole
 \cite{GroPer:} in which the
 internal direction is a timelike circle, and therefore exists in a
 spacetime of signature (3,2). Interestingly, it seems that there are no
 constraints on the energy of the solution. In general, for any
 supersymmetric theory
 with two times,  there is no intrinsic concept of energy or
length in the same way as for $T=1$ supersymmetry. One corollary of
 this is that  we
do not need concern ourselves with the problems that arise if one tries
to define a naive $T=2$ quantum supersymmetric system.  The resolution
of the quantisation procedure remains mysterious, although it is
likely to be formally possible with  path integral formulation. This
 form of quantisation has already been  carried out for the
 (2,2)-string \cite{BisLec:}\footnote{It is worth noting that a nice
 property of   path integrals in a $T=2$ spacetime  is that the
 action is real, as opposed to the imaginary quantity which occurs in a
 Minkowski signature theory.}.

\section{$T=1$ SUSY and $T=2$ SUSY}
We now consider the basic supersymmetry theories containing BPS
solutions with one or two timelike directions. 
If $T=1$ then due  to the positive energy condition the simplest possible
non-trivial anticommutator  is 
\be\label{f1}
\{Q^\alpha,Q^\beta\}_{T=1}=\left(C\Gamma^\mu\right)^{\alpha\beta}P_\mu\,.
\ee
The simplest superalgebra with a non-trivial bosonic sector which includes the Lorentz rotations consistent
with this anticommutator is in fact the standard super \po
algebra. This has the non-zero terms additional to (\ref{f1}) 
\ben\label{sio1terms}
[M_{\mu\nu},M_{\rho\sigma}]&=& 
M_{\nu\sigma}\eta_{\mu\rho} +M_{\mu\rho}\eta_{\nu\sigma}-M_{\nu\rho}\eta_{\sigma\mu}- M_{\sigma\mu}\eta_{\nu\rho} \nonumber \\
\left[M_{\mu\nu},P_\rho\right]&=&P_\mu\eta_{\nu\rho}-P_\nu\eta_{\mu\rho}\nonumber \\
\left[Q^{\alpha},M_{\mu\nu}\right]&=&\frac{1}{2}{\left(\Gamma_{\mu\nu}\right)^\alpha}_\beta
Q^\beta\,.
\een
In analogy with choosing $Z^\mu\rightarrow P^\mu$, it seems natural to
make the identification $Z^{\mu\nu}\rightarrow M^{\mu\nu}$. This was
the approach discussed in \cite{HewPer:22brane}. If we adopt this
identification then the  two-time commutator  analogous to (\ref{f1})
is 
\be\label{f2}
\{Q^\alpha,Q^\beta\}_{T=2}=\left(C\Gamma^{\mu\nu}\right)^{\alpha\beta}M_{\mu\nu}\,.
\ee
Since only the rotation  generators appear in this expression, it has
a simpler extension than  the \po 
algebra: The full algebra, called \siot is given by
(\ref{f2}) with the additional terms 
\ben\label{adsterms}
[M_{\mu\nu},M_{\rho\sigma}]&=& 
M_{\nu\sigma}\eta_{\mu\rho} +M_{\mu\rho}\eta_{\nu\sigma}-M_{\nu\rho}\eta_{\sigma\mu}- M_{\sigma\mu}\eta_{\nu\rho} \nonumber \\
\left[Q^{\alpha},M_{\mu\nu}\right]&=&\frac{1}{2}{\left(\Gamma_{\mu\nu}\right)^\alpha}_\beta
Q^\beta\,.
\een
It is a simple matter to check that a representation of the
supersymmetric \siot system is given by 
\ben\label{adsrep}
M_{\mu\nu}&=&{\widetilde{M}}_{\mu\nu}+\frac{1}{4}\theta_\alpha\left(\Gamma_{\mu\nu}\right)^{\alpha\beta}\diff_\beta\nonumber\\
Q^\alpha&=&{\diff\theta^\alpha}-\frac{1}{2}{\left(\Gamma^{\mu\nu}\right)^{\alpha}}_\beta
Q^\beta {M}_{\mu\nu}\,,
\een
where ${\widetilde M}$ is the orbital part of the angular momentum
operator, defined by
\be\label{tt2a}
{\widetilde M}_{\mu\nu}=X_{[\mu}\diff_{\nu]}\,.
\ee
Note that for consistency of this algebra  we require that 
\be
(\Gamma^{\hat{\mu}\hat{\nu}})^{{(}\alpha\beta}(\Gamma_{\hat{\mu}\hat{\nu}})^{\gamma\delta{)}}=0\,,
\ee
where the indices $(\hat{\mu}\hat{\nu})$ occur whenever
$M_{\hat{\mu}\hat{\nu}}$ is non-zero. This is easily seen to be the
case  by
evaluation of  the $\{QQQ\}$ super Jacobi identity. 
 Of course, if we choose
$\{Q,Q\}$ to be {\it centrally} extended with a $Z^{\mu\nu}$ term, then the
algebra is always consistent. For the purpose of the results obtained
in this paper, this distinction is of no consequence.

\subsection{BPS solutions}
We now search for BPS solutions of the basic  $T=1$ and $T=2$
systems with anticommutators (\ref{f1}) and  (\ref{f2}). 
For $T=1$ we note that  in any dimension the determinant of the
anticommutator (\ref{f1}) is zero 
if and only if  $P^\mu P_\mu=0$. The positive energy condition
requires that $P_0>0$,
hence the solution is the massless supersymmetric  Brinkmann wave
\cite{Kal:dualwaves}. 
Although there is a vast literature on the BPS solutions to (\ref{f1})
with central extensions in dimensions ten and eleven,  
many may be derived from this plane wave solution  by using $S$-,
$T$-dualities  and dimensional reductions \cite{Gau:intersecting}. 
We may attempt to find analogies when $T=2$. 
It is, unfortunately, more difficult to analyse the solution to
$\det\{Q^\alpha,Q^\beta\}_{T=2}=0$.

 We proceed by considering canonical forms for the components of the
 rotation generator $M_{\mu\nu}$, which may be obtained from more
 general cases by the action  of Lorentz transformations. To begin a
 partial classification  we first note that $S_{\mu\nu}\equiv
 M_{\mu\rho}M_{\nu\sigma}\eta^{\rho\sigma}$ is a real symmetric
 matrix, whose eigenvalues must consist of repeated pairs, since the
 eigenvalues of an antisymmetric matrix occur in pairs of opposite
 sign.  
 The eigenvectors of the symmetric matrix may be spacelike, timelike
 or null. A generalisation of the classification of symmetric matrices
 in signature (3,1) \cite{HawEll:} to signature (2,2) shows us that
 the only cases consistent with the specific form of $S_{\mu\nu}$ as
 the square of the rotation matrix  
are those for which  $S_{\mu\nu}$ is diagonalisable in a orthonormal
 basis of eigenvectors. The corresponding  canonical forms of
 $M_{\mu\nu}$ are as follows 
\be\label{canonical22} 
\bordermatrix{&\scriptstyle{1}&\scriptstyle{2}&\scriptstyle{0}&\scriptstyle{\tilde{0}}\cr
 &0&a&b&c\cr &-a&0&\mp c&\pm b\cr &-b&\pm c&0&\pm a\cr &-c&\mp b&\mp
 a&0}\,, 
\pmatrix{0&a&0&0\cr -a&0&0&0\cr 0&0&0&f\cr 0&0&-f&0}\,,
\pmatrix{0&0&0&c\cr 0&0&d&0\cr 0&-d&0&0\cr -c&0&0&0}\,,
\pmatrix{0&0&b&0\cr 0&0&0&e\cr -b&0&0&0\cr 0&-e&0&0}\,.
\ee
To compare this result with the standard scenarios, we note that
in signature (3,1) we only obtain matrices of the form of the
 last three matrices of (\ref{canonical22}), whereas in 
 a positive definite   space antisymmetric matrices are block
 diagonal, with  off diagonal blocks. Using the basic blocks
 (\ref{canonical22}) allows 
 us to make some 
 start at the  analysis in (10,2).   

Suppose first that the angular momentum lies in a (2+2) dimensional
subspace. The last two forms of (\ref{canonical22}) are unsuitable for
two-timing BPS states, since the $0\tilde{0}$ component of angular momentum is
zero.  
We thus have two inequivalent choices to investigate, for which we
may calculate the determinant explicitly.  
In any theory with both a timelike and spacelike direction we may
represent the gamma matrices as the real matrices
\be\label{realrep}
\pmatrix{\Delta_p&0\cr0&-\Delta_p}\,,\pmatrix{0&1\cr -1&0}\,,\pmatrix{0&1\cr1&0}\,,
\ee
where the $\Delta_p$ are the gamma matrices of the $(S-1,T-1)$
Clifford algebra \footnote{Since the reality properties of C(S,T) and
$C(S-1,T-1)$ are the same, we may choose the $\Delta_p$ to be real for
a Majorana representation of C(S,T).}.

Recall that a BPS configuration is a set of fields such that the
determinant of the anticommutator $\{Q^\alpha,Q^\beta\}$ is
zero. Equivalently we may search for zero eigenvalues of the
matrix. We now do this for the possible
canonical forms  (\ref{canonical22}):
\begin{enumerate}
\item
Employing the spin-space reduction (\ref{realrep}) twice, we see that
the characteristic 
equation of the first matrix of (\ref{canonical22}) is given by 
\be\label{chareqn}
\lambda^2\left(\lambda^2-2a\lambda+(a^2-b^2-c^2)\right)=0\,.
\ee
We therefore have a supersymmetric solution with two zero eigenvalues
for {\it any} values  of $a,b$ and $c$. In the special case for which
we have $a^2=b^2+c^2$ there  is extra supersymmetry due to  an
additional zero eigenvalue. This additional condition on $a,b$ and $c$
is satisfied if and only if $M^2\equiv M_{\mu\nu}M^{\mu\nu}=0$. By
squaring the RHS of the expression  
\be\label{solution2}
\{Q,Q\}=\Gamma^{0\tilde{0}}\left(a(\Gamma^{0\tilde{0}}\pm\Gamma^{12})+b(\Gamma^{01}\pm\Gamma^{\tilde{0}2})+c(\Gamma^{02}\mp\Gamma^{1\tilde{0}})\right)\,,
\ee
we see that the $M^2=0$ cases are precisely those for which the right
hand side of the anticommutator is proportional to a projection
operator.  
This is a pleasing result, and these solutions are naturally to be
interpreted as the two-time
analogues of the Brinkmann waves. The relation may be seen thus 
\be
(T=1)\quad P_0>0\,,\quad P_\mu P^\mu=0\longleftrightarrow
M_{\mu\nu}M^{\mu\nu}=0\,,\quad M_{0\tilde{0}}>0
\quad (T=2)
\ee

\item
For the second possible canonical form of the matrix we find that 
\be
\{Q,Q\}= 2f1+2a\Gamma^{0\tilde{0}12}\,,
\ee
which corresponds to a BPS state if and only if $a=|f|$, in which case
$M^2=4a^2$. This solution naturally represents a {\it (2+2)-brane} in the 1-2
direction. A Green-Schwarz formulation of the (2+2)-brane is provided
in \cite{HewPer:22brane}. 
\end{enumerate}

In the next section we use these ideas to study the BPS configurations
to the algebra of
relevance to F-theory.

\section{F-branes}
It is well known that properties of a locally supersymmetric theory may
be deduced from a study of the corresponding rigid supersymmetry
theory: Many properties of M-theory may be deduced from the study of
the eleven dimensional supersymmetry algebra which has the anticommutator
\be\label{11d}
\{Q^\alpha,Q^\beta\}=\left(C\Gamma^\mu\right)^{\alpha\beta}P_\mu+\left(C\Gamma^{\mu\nu}\right)^{\alpha\beta}Z_{\mu\nu}+(C\Gamma^{\mu\nu\rho\sigma\lambda})^{\alpha\beta}Z_{\mu\nu\rho\sigma\lambda}\,.
\ee
From the viewpoint of supersymmetric $T=1$ physics, 
this algebra is often  considered to be fundamental,
since it reduces to all the supersymmetry algebras of the different
types of superstring theories in {\it nine} dimensions,
either directly or by appealing to T-duality \cite{Tow:fourlectures};
consequently all the
information about supergravity BPS states may be obtained by studying
the solutions to this algebra. However, we need not stop at eleven
dimensions: It is well known that 
the superalgebra (\ref{11d}) arises as a Wigner-Inounou contraction of
an ortho-symplectic algebra in twelve dimensions
\cite{HewPer:22brane,Tow:democracy,HolPro:}, which is given by
\ben\label{12d}
[M_{\mu\nu},M_{\rho\sigma}]&=& 
M_{\nu\sigma}\eta_{\mu\rho}
+M_{\mu\rho}\eta_{\nu\sigma}-M_{\nu\rho}\eta_{\sigma\mu}-
M_{\sigma\mu}\eta_{\nu\rho} \nonumber \\
\{Q^\alpha,Q^\beta\}_{(10,2)}&=&\left(C\Gamma^{\mu\nu}\right)^{\alpha\beta}Z_{\mu\nu}+\left(C\Gamma^{\mu\nu\rho\sigma\lambda\delta}\right)^{\alpha\beta}Z^+_{\mu\nu\rho\sigma\lambda\delta}\,.
\een
The $Q^\alpha$ in this expression are taken to be 32 component
Majorana-Weyl spinors, and the
central six-form field is self-dual. This is a maximal supersymmetry algebra
in the sense that it has (528+528) degrees of freedom, as does
(\ref{11d}); yet it has the advantage that it reproduces both the IIA
 and  IIB theories directly in ten dimensions \cite{HewPer:22brane}. 
We call this algebra the F-algebra, and there are clearly good reasons
to believe that it could be considered to be Fundamental. 

 When the F-algebra is reduced, 
 the momentum operator $P^\mu$ is obtained 
from the Kaluza-Klein reduction of the
operator $Z_{\mu\nu}$. If we compactify on the
$\tilde{0}$-direction then the momentum is simply
$P_{\mu}=Z_{\tilde{0}\mu}$ and the energy of the $T=1$ system is 
given by the component $J=Z_{0\tilde{0}}\rightarrow P_0=E$. Hence the
two positivity 
conditions $J\geq 0$ and $E\geq 0$ are consistent with each other.

\subsection{Fundamental branes in (10,2)}
We now turn to the question of the branes which appear in
F-theory. In analogy with results from M-theory, we anticipate that  an
algebraic understanding of  
the solutions 
of (\ref{12d}) will enlighten us as to the nature of F-theory. 
In what follows we
define the two timelike dimensions to be labelled by $0,{\tilde{0}}$.
 It is well worth noting that since the F-algebra is   of the form 
\be
\{Q,Q\}=(C\Gamma^{\mu_1\mu_2})Z_{\mu\nu}+(C\Gamma^{\mu_1\dots\mu_6})Z_{\mu_1\dots\mu_6}\,,
\ee
we would perhaps expect to see 2-branes and 6-branes
\cite{Duf:mtheory}, in accordance with the $T=1$ theory. As we
shall see, in this
picture we
obtain 3-branes and 7-branes because the charge conjugation matrix is
now a product of {\it two} gamma matrices: In general for a
two-timing theory the appearance of a $p$-form charge in the algebra
implies the existence of a (p+1)-brane.

\subsubsection{Case 1: $Z^{\mu\nu\rho\sigma\lambda\delta}=0$}
We begin by setting all of the $Z^{\mu\nu\rho\sigma\lambda\delta}$
terms to zero, and  search for vanishing determinant configurations of
\be\label{ac1}
\{Q^\alpha,Q^\beta\}_{(10,2)}=\left(\Gamma^{0{\tilde{0}}}\Gamma^{\mu\nu}\right)^{\alpha\beta}Z_{\mu\nu}\,.
\ee
 We find, of course, all the (2+2)-dimensional solutions given in the previous
sections, but there also exist solutions such as 
\be\label{solution1}
Z_{0\tilde{0}}=J, Z_{0p}=J_p\, \mbox{ with }
\sum_{p=1}^{10}(J_p)^2=J^2\,.
\ee
All of the solutions we have been able to find, excepting the
(2+2)-brane, have $Z_{\mu\nu}M^{\mu\nu}=0$.

\subsubsection{Case 2: $Z^{\mu\nu\rho\sigma\lambda\delta}\neq 0$}
We now   include the central six-form charge terms into the algebra and
study the cases  
with only the necessary $Z_{0\tilde{0}}=J$ part of the rotation non-zero
\be\label{ac2}
\{Q^\alpha,Q^\beta\}=J(1)^{\alpha\beta}+\left(\Gamma^{0\tilde{0}}\Gamma^{\mu\nu\rho\sigma\lambda\delta}\right)^{\alpha\beta}Z_{\mu\nu\rho\sigma\lambda\delta}\,.
\ee
In order to find some determinant zero solutions, we look for cases
for which the right hand side  is a projection 
operator. In theories with one timelike variable, we are interested in
configurations which have compact transverse group. From the algebraic
point of view this requires that all the indices on the form
$Z_{\mu_1\dots\mu_p}$ corresponding to the $p$-brane are spacelike
variables. We find that there are two possible BPS configurations in
(10,2) consistent with this requirement after reduction to a $T=1$ theory
\ben
Z_{pqrstu}&=&J,\hskip1cm\mbox{all other terms zero,}\nonumber\\
Z_{\tilde{0}pqrst}&=&J,\hskip1cm\mbox{all other terms zero}\,.
\een
The first of these is easily interpreted as being a (6+2)-brane in
the $pqrstu$ plane, and  the second corresponds to a (6+2)-brane in the
$pqrst$ plane, with the $\tilde{0}$ direction in the  brane
worldvolume. Notice that algebraically this second configuration is
equivalent to that of a BPS 5-brane solution in a $T=1$ theory as follows
\be
\{Q,Q\}=J\left(1+\Gamma^{0\tilde{0}}\Gamma^{\tilde{0}pqrst}\right)=J\left(1-\Gamma^{0}\Gamma^{pqrst}\right)\,.
\ee

In general, the factor in front of the
$Z$ terms need not be equal to $J$ for a general, non-BPS state; for
positivity we just require that $J\geq M$. The BPS states are obtained
when the bound is saturated. Note that in order to conform to standard
notation, we shall refer to a brane which couples to a six-form
charge as a seven brane, since 7+1=6+2. In this way we interpret the
(5+1) brane as a seven brane for which one of the indices of the
six-form charge is timelike.

\subsection{Fundamental branes and magnetic branes}
In traditional supersymmetry, a fundamental brane is an electrically
charged object, in the sense that there is a point singularity
corresponding to the source. The total charge is then defined to be
\be
Z^{\mu_1\dots\mu_p}=\int_{S^{D-d-1}} \star F_{d+1}\,,
\ee
where $F$ is the field strength to which the brane with $d=p+t$ extended
dimensions couples to. We may thus suppose that for each charge in the
supersymmetry algebra there exists an associated fundamental brane. 
In our case these are the  BPS 3- and 7-branes.  By writing the action for the
coupling to the brane of the field strength $F$  in terms of  the \po
dual of a $(D-d-1)$ index dual field strength, we may define the
magnetic partner of  the original electrically
charged brane. Such an 
argument is signature independent and thus we may define magnetic
duals in a $T=2$ theory. This gives us the pairings
\be
(2+2)\longleftrightarrow(4+2)\hskip1cm \mbox{and}\hskip1cm (6+2)\longleftrightarrow
(0+2)\,.
\ee 
The (4+2)-brane is self explanatory; the (0+2)-brane corresponds
to the two-timing version of a particle, for which the worldsheet is a
timelike plane.

\subsection{Worldvolume theories}
We now turn to the question of the  worldvolume content  of the
F-branes. The BPS condition is a question purely about the rigid
{\it spacetime} 
supersymmetry of the theory, whereas 
 we presumably  require that the
fields on the {\it worldsheet} form a local supersymmetry theory. A simple
way to deduce the contents of the multiplet is to note that for a
supersymmetry theory on a worldsheet of dimension greater than one the
bosonic and  fermionic degrees of freedom must match up. This is true
even  if the commutator $\{Q,Q\}$ does not generate momentum. 
To see that this is indeed the case, we note that the representation
space of a supersymmetry theory may be considered to be the disjoint
union of a bosonic subspace ${\cal{B}}$ and a fermionic subspace
${\cal{F}}$. The  supersymmetry generator $Q$ by definition maps
${\cal{F}}$ onto a proper subspace of ${\cal{B}}$ and vice
versa. However, the product of two supersymmetry transformations is a
bosonic operator,  the particular operator being dependent on the
theory in question: For the \po  supersymmetry it is a momentum
generator $P$, whereas more generally it is some other bosonic
operator. Each of these possible operators  clearly map the representation
space onto itself. For this reason the action of the supersymmetry generator
must provide a one to one mapping from ${\cal{B}}$ to ${\cal{F}}$, and
we therefore have the result that the worldsheet supermultiplet must have
{\it equal numbers of bosons and fermions}, for both the eleven
dimensional superalgebra and the
F-algebra.

So, in order to investigate the matching of the bosons with the fermions  we
need to determine the 
the on-shell degrees of freedom of tensorial operators in spaces with
two timelike directions. To evaluate the answer we see that 
in a quantum
theory each ghost field absorbs one spacelike and one timelike mode,
leading to a $T=1$ vector index possessing $D-2$ degrees of 
freedom. A theory with two times ought to have twice as many ghosts as
a Minkowskian theory \cite{BarGibPerPopRub:}. 
Therefore in a $T=2$ theory a  vector index has $D-4$ degrees of
freedom in the analogue of the light 
cone gauge. Such a statement leads to very  interesting results in the
context of $N=2$ strings, which have worldsheet theories which are 
effectively 2-complex dimensional \cite{OogVaf:geometry}.  

We may now apply the matching conditions to the F-branes to find the
following worldsheet field contents 
\be\label{doftable}
\begin{array}{c|c|c}
(s+t)&n_f-n_b&{\mbox{Multiplet}}\\\hline
(0+2)&-2&-\\
(2+2)&0&\{X^{\mu_1}\dots X^{\mu_8};\theta^{\alpha}\}\\
(4+2)&2&\{X^{\mu_1}\dots X^{\mu_6},A^\mu;\theta^{\alpha}\}\\
(6+2)&4&\{X^{\mu_1}\dots X^{\mu_4},H^{\mu\nu\rho}\sim A^\mu;\theta^{\alpha}\}\\
\end{array}
\ee
Notice that for a worldsheet of dimension (6+2) a three-form field and
a one-form field have the same degrees of freedom. 
For the (4+2) and the (6+2) branes we may also consider the dual
worldsheets fields, to which other F-branes may couple. The field
strengths associated with the worldsheet potentials are denoted by $F$. 
\be\label{dualtable}
\begin{array}{c|c|c|c}
\mbox{Brane}&F&\tilde{F}&\mbox{Coupling}\\\hline
(4+2)&F^{\mu\nu}&F^{\mu\nu\rho\sigma}&(2+2)-\mbox{brane}\\\hline
(6+2)&F^{\mu\nu\rho\sigma}\,, F^{\mu\nu}&F^{\mu\nu\rho\sigma}\,, F^{\mu\nu\rho\sigma\delta\lambda}&(2+2)/(4+2)-\mbox{branes}
\end{array}
\ee

\subsection{Intersection rules}
Now that  we have defined the basic fundamental and magnetic branes
associated with the twelve dimensional F-superalgebra we may try to
mimic the theory of intersecting branes in Minkowski spacetimes
\cite{Gau:intersecting}. This approach has proven to be very useful in
understanding the restrictions that supersymmetry places on supergravity brane
configurations. 
We therefore  wish to investigate briefly the intersection rules for the
fundamental (2+2)- and (6+2)-branes.   There are several
possibilities, which we detail as follows:

\begin{enumerate}
\item
For the algebra (\ref{ac1})  the projectors corresponding to  two
(2+2)-branes are given by  
$\frac{1}{2}(1+\Gamma^{0\tilde{0}pq})$ and
$\frac{1}{2}(1+\Gamma^{0\tilde{0}rs})$. For an intersecting configuration to be
possible we must have that these projectors commute with each
other, so that their product is also a projector.  Assuming that $pq$ and
$rs$ are not the same, the projectors 
commute provided that the $p,q,r,s$ are all distinct. Thus  two
(2+2)-branes may intersect on a (0+2)-brane. This may be viewed
schematically as
\be
\begin{array}{|c|c|c|}\hline
0\tilde{0}&pq&\\
0\tilde{0}&&rs\\\hline
\end{array}
\ee
\item
The projectors corresponding to  two (6+2)-branes may be taken to be  given by 
{$\frac{1}{2}(1+\Gamma^{0\tilde{0}pqrstu})$} and
$\frac{1}{2}(1+\Gamma^{0\tilde{0}pqwxyz})$. For these two projectors
to commute, an even number of the spatial indices must match up. Clearly in
twelve dimensions this number must be at least two. 
The results are that the two (6+2)-branes may intersect on (2+2)- and
(4+2)-branes, as follows
\be
\begin{array}{|c|c|c|c|}\hline
0\tilde{0}&pq&rstu&\\
0\tilde{0}&pq&&vwxy\\\hline
\end{array}
\ee
\be
\begin{array}{|c|c|c|c|}\hline
0\tilde{0}&pqrs&tu&\\
0\tilde{0}&pqrs&&vw\\\hline
\end{array}
\ee

These intersections are in agreement with the possible worldsheet
couplings of the branes given in (\ref{dualtable}).
\item
If we choose one of the indices on the six-form to be timelike then it
transpires that the the solution is still supersymmetric. In this case
the intersection relations `contract' and we obtain (5+1)-brane intersections
\be
\begin{array}{|c|c|c|c|}\hline
0&q&rstu&\\
0&q&&vwxy\\\hline
\end{array}
\ee
\be
\begin{array}{|c|c|c|c|}\hline
0&qrs&tu&\\
0&qrs&&vw\\\hline
\end{array}
\ee

\noindent As we shall see in the following section, these 5-brane
solutions may be considered to
be the trivial lifts to twelve dimensions of  M-theory  brane
configurations. 
\end{enumerate}

\subsection{Reduction to lower dimensions}
We now consider the reduction of the F-branes in a purely
super-algebraic way. 
To proceed we need to consider carefully the effects of the
reduction on the supersymmetry theory.
We may suppose that the BPS F-brane configurations may be written as  
\be
\{Q,Q\}={\cal{P}}\,,
\ee
for some  projection operator ${\cal{P}}$. To obtain the lower
dimensional form of this anticommutator in ten and eleven dimensions
we must act on the spinors 
with the  projectors 
${\cal{P}}_{10}=\frac{1}{2}(1+\Gamma^0\dots\Gamma^9)$ and 
${\cal{P}}_{11}=\frac{1}{2}(1+\Gamma^0\dots\Gamma^9\Gamma^{11})$
respectively. There are two distinct
types of supersymmetry theory which may arise after  the
compactification. Firstly, ${\cal{P}}$ could be mapped onto  another
projector in the lower dimension. In this 
case the 
surviving terms would correspond to a BPS configuration in either 
type IIB  or M-theory. Secondly, there is the 
possibility that the RHS of the algebra would become zero in the
reduced theory. This would correspond to a brane propagating in a
trivially realised supersymmetry theory, for which the superspace is flat
with no torsion. The role of these types of superspaces (which have
been investigated previously \cite{HewPer:22brane,Hew:generalised}) 
in M-theory 
has not yet been determined. We shall therefore concentrate on the
solutions which are non-trivially realised in lower dimensions, and 
investigate the various possibilities in turn. 

\subsubsection{Brinkmann wave analogues}
We first look at solutions which  have only $Z_{\mu\nu}$ charges, with
$Z_{\mu\nu}Z^{\mu\nu}=0$; the Brinkmann wave analogues.  An example which
qualitatively covers the features of many higher dimensional solutions
is the four dimensional example (\ref{solution2}), given by 
\be\label{solution3}
\{Q,Q\}=\Gamma^{0\tilde{0}}\left(a(\Gamma^{0\tilde{0}}\pm\Gamma^{1s})+b(\Gamma^{01}\pm\Gamma^{\tilde{0}s})+c(\Gamma^{0s}\mp\Gamma^{1\tilde{0}})\right)\,,
\ee
where the index $s$ may take any spacelike value. We thus  have
$Z_{\tilde{0}0}=\pm Z_{1s}$ etc. For the BPS states we impose the
condition that $a^2=b^2+c^2$.  
To obtain a ten dimensional theory we reduce on the torus with
coordinates $\tilde{0}$ and $\tilde{1}$. There are two distinct cases
\begin{itemize}
\item
If $s=1,\dots, 9$ then the surviving terms are $Z_{\tilde{0}0}$,
$Z_{\tilde{0}s}$ and $Z_{\tilde{0}1}$. In the reduced theory the
anticommutator then becomes 
\be
\{Q,Q\}=P_0+(C\gamma^s)P_s\pm (C\gamma^1)P_1\,,
\ee
where all the quantities are now ten dimensional. Squaring this
operator, we find that $(P_0)^2=(P_1)^2+(P_s)^2$, which therefore
corresponds to a Brinkmann wave with $P^2=0$.  
\item
If $s=\tilde{1}$ then the quantities which remain after the projection
are $Z_{\tilde{0}0}=\pm Z_{\tilde{1}1}$ and
$Z_{\tilde{0}1}=Z_{\tilde{1}{0}}$  
\be
\{Q,Q\}=P_0(1\pm(C\gamma^1))\,,
\ee
which is again a massless plane wave. 
\end{itemize}
The reduction of the solutions (\ref{solution3}) to eleven dimensions
on the timelike coordinate $\tilde{0}$ provides us with only one
massless plane wave 
\be
\{Q,Q\}=P_0\pm(P_1(C\Gamma^1)+P_2(C\Gamma^2))\,.
\ee
An example not covered by these cases is (\ref{solution1}), for which
$Z_{0\tilde{0}}=J$ and $Z_{0s}=J_s\,:\,\sum_{s=1}^{1}(J_s)^2=J^2$. If
we reduce this  solution to  eleven dimensions  
then the only surviving generator in the algebra is the
$Z_{0\tilde{0}}$ component, and we therefore find a massive
particle. A toroidal compactification provides us with another such
object in ten dimensions.

\subsubsection{3-branes and 7-branes}
The next solution to consider is the three-brane. The
reduction of the Green-Schwarz (2+2)-brane was discussed in 
\cite{HewPer:22brane}, 
and was shown to produce the type IIB string and the M-2-brane. 
We are consequently left to analyse  the seven-branes, which
have non-vanishing six-form charges. We look at the simplest examples,
of the form 
\be
\{Q^\alpha,Q^\beta\}=J(1)^{\alpha\beta}+\left(\Gamma^{0\tilde{0}}\Gamma^{\mu\nu\rho\sigma\lambda\delta}\right)^{\alpha\beta}Z_{\mu\nu\rho\sigma\lambda\delta}\,,
\ee
with a single non-zero six-form component. We are interested in the
cases for which
$\widetilde{{\cal{P}}}_{10/11}(C\Gamma^{\mu\nu\rho\sigma\lambda\delta}){\cal{P}}_{10/11}$
is non-zero. If we reduce to eleven dimensions then  the $Z_{pqrtsu}$
case is projected to zero, whereas for
$Z_{\mu\nu\rho\sigma\lambda\delta}\equiv Z_{\tilde{0}pqrst}=J$ we
obtain a BPS saturated five brane 
\be
\{Q,Q\}_{11}=J(1+(C\Gamma^{pqrst})Z_{pqrst})\,. 
\ee
If we perform the double dimensional reduction of the theory on a
torus down to the IIB theory then the converse  situation holds:  It is simple
to see that the $Z_{\tilde{0}\tilde{1}pqrs}$ term 
vanishes, whereas the term corresponding to a (6+2)-brane,
$Z_{\tilde{1}pqrst}$, takes us to a five-brane in the IIB theory\footnote{By
altering the parameters of the torus we reduce upon is seems possible
in principle to generate the complete $SL(2,\zz)$ 5-brane multiplet,
in the same manner that the $SL(2,\zz)$ invariance of the type IIB
string is explained by F-theory}. 
This is rather interesting: Although we have seven branes in the
twelve dimensional theory, we can only produce five-branes under
dimensional reduction! This is pleasing because there are no
fundamental or solitonic six-branes in the spectrum of M-theory. 

We should now inquire as to the effect of the reduction on the
worldvolume fields of the seven-branes. It is sufficient to note that
the supersymmetry theory on the brane will be reduced onto another
supersymmetry theory on the new brane in a lower dimension. Since the
worldvolume content of the five-branes in ten and eleven dimensions
may be deduced from the considerations of a six-dimensional
supersymmetry theory, we may be sure that the worldvolume content of
the 12-dimensional seven branes will be mapped onto the correct
supersymmetry theories for the well known Minkowski signature
five-branes. To summarise, we have the correspondences
\be
\begin{array}{c|c|c}
F&M&IIB\\\hline
7&5&5\\
3&2&1
\end{array}
\ee

\subsubsection{Intersection rules reduction}
We now finally wish to reduce the intersection rules we obtained
previously. It is important to consider the reduction to M-theory,
since the M-branes may be used to reproduce all of the branes in
string theory via dimensional reduction and the use of dualities
\cite{Gau:intersecting}. As a corollary, all of the intersection rules
in the type II theories descend from those in eleven dimensions. 
We find that the 
reduction of the intersecting F-7-brane configurations on a timelike
circle provides us with two 
M-5-branes intersecting on either 3-branes or strings. Reduction of
the intersecting F-3-branes gives two M-2-branes intersecting on a
0-brane. These are precisely the basic M-brane configurations from
which all the supergravity brane configurations may be deduced, hence
we have complete consistency between the two scenarios. 

\section{Conclusion}
We have discussed the basic algebraic consequences of BPS states in 
 supersymmetry theories with two timelike directions. In theories with a
single time variable,  every low energy BPS configuration must have a
positive energy. Quite independently of any one-time considerations,
we showed that a supersymmetric BPS state in {\it any} four or greater
 dimensional 
theory  with two times must have a `positive angular momentum in the
 timelike plane': $Z_{0\tilde{0}}=J>0$. We then studied the algebraic
 restrictions on the 
$p$-branes arising from the supersymmetry algebra in such
 theories. This type of procedure has 
proven to be very useful in M-theory, since each BPS solution of the
supersymmetry algebra corresponds to some  supergravity configuration 
which is the low energy limit of an excitation of the underlying
quantum theory. 
Clearly, there is a possibility that the M-theory structure may
derive from some theory with ten spacelike and two timelike
directions.  If it
does, then  this  theory 
is sure to be one  of supersymmetric brane configurations;
otherwise, there is very little which may be said with certainty. For
example, it is even unclear as to whether or not the twelve
 dimensional theory would
reduce to the Einstein-Hilbert action in the low energy limit: As we
have shown in this paper, there is no fundamental notion of energy
in a theory with two timelike directions. The philosophy of the work
presented here is that quite generally  any supersymmetric theory must
in a fundamental way have behaviour governed by the tangent space superalgebra
structure.
For this reason we 
studied  the F-algebra, which is the algebra most likely to be
 relevant to M-theory. The basic BPS configurations of the F-algebra
 are  3-branes and 7-branes, along with magnetic dual
partners. This resulting brane 
structure when reduced to eleven dimension reproduces precisely the
brane structure of eleven dimensional superalgebra; when reduced on a
null-torus we reproduce the type IIB string and fivebrane
 solutions. In addition to these standard backgrounds there exists the
 possibility of a compactification down to a simple supersymmetry
 sector of the ten and eleven dimensional theories, in which the 
 anticommutator of the supersymmetry generators 
 vanish.

Although
the discussion we have presented here is completely classical, it is
reassuring to note that if we 
allow ourselves to use the known duality relations in lower
dimensions,  
there is a direct correspondence between the  F-branes and those
of the Minkowski signature theories. From this structure, it should
now be possible to start to lift the M-branes to the known BPS
F-branes. Hence, by using this algebraic correspondence 
we may begin to understand more of the meaning of F-theory.  

\section{Acknowledgements}
The author would like to thank the EPSRC and Queens' College,
Cambridge for their financial support.

\end{document}